\documentclass[prd,twocolumn,aps,superscriptaddress,showpacs]{revtex4}
\usepackage{amssymb}
\usepackage{amsmath,bm}
\usepackage{graphicx}
\usepackage[normalem]{ulem}
\usepackage[dvips]{color}
\usepackage{subfigure}
\newcommand{\be}{\begin{equation}}

\newcommand{\ee}{\end{equation}}
\newcommand{\bea}{\begin{eqnarray}}
\newcommand{\eea}{\end{eqnarray}}

\renewcommand\sout{\bgroup \color{red} \ULdepth=-.5ex \ULset}

\begin{document}

\title{Quark magnetar in three-flavor Nambu--Jona-Lasinio model
   with vector interaction and magnetized gluon potential}
\author{Peng-Cheng Chu}
\email{kyois@sjtu.edu.cn}
\affiliation{Department of Physics and Astronomy and Shanghai Key Laboratory for Particle
Physics and Cosmology, Shanghai Jiao Tong University, Shanghai 200240, China}
\affiliation{Institute of High Energy Physics, Chinese Academy of Sciences, Beijing 100049, China}
\author{Xin Wang}
\email{xinwang@sjtu.edu.cn}
\affiliation{Department of Physics and Astronomy and Shanghai Key Laboratory for Particle
Physics and Cosmology, Shanghai Jiao Tong University, Shanghai 200240, China}
\author{Lie-Wen Chen}
\email{lwchen@sjtu.edu.cn}
\affiliation{Department of Physics and Astronomy and Shanghai Key Laboratory for Particle
Physics and Cosmology, Shanghai Jiao Tong University, Shanghai 200240, China}
\affiliation{Center of Theoretical Nuclear Physics, National Laboratory of Heavy-Ion
Accelerator, Lanzhou, 730000, China}
\author{Mei Huang}
\email{huangm@mail.ihep.ac.cn}
\affiliation{Institute of High Energy Physics, Chinese Academy of Sciences, Beijing 100049, China}
\affiliation{Theoretical Physics Center for Science Facilities, Chinese Academy of Sciences,
Beijing 100049, China}

\begin{abstract}
We investigate properties of strange quark matter in the framework of SU(3)
Nambu--Jona-Lasinio(NJL) model with vector interaction under strong magnetic
fields. The effects of vector-isoscalar and vector-isovector interactions on
the equation of state of strange quark matter are investigated,
and it is found that the equation of state is not sensitive to the vector-isovector
interaction, however, a repulsive interaction in the vector-isoscalar channel gives
a stiffer equation of state for cold dense quark matter. In the presence of magnetic
field, gluons will be magnetized via quark loops, and the contribution from
magnetized gluons to the equation of state is also estimated. The sound velocity
square is a quantity to measure the hardness or softness of dense quark matter,
and in the NJL model without vector interaction at zero magnetic field the sound
velocity square is always less than 1/3. It is found that a repulsive vector-isoscalar
interaction and a positive pressure contribution from magnetized gluons can
enhance the sound velocity square, which can even reach 1. To construct quark magnetars
under strong magnetic fields, we consider anisotropic pressures and use a
density-dependent magnetic field profile to mimic the magnetic field distribution
in a quark star. We also analyze the parameter region for the magnitude of
vector-isoscalar interaction and the contribution from magnetized gluons in order to
produce 2 solar mass quark magnetars.
\end{abstract}

\maketitle

\section{Introduction}

Investigating properties of strong interaction matter is one of the main
topics of quantum chromodynamics (QCD). It is believed that there will be
deconfinement phase transition from hadronic matter to quark-gluon plasma (QGP)
at sufficiently high temperatures and from nuclear matter to quark matter
(or color superconductor) at high baryon densities. The hot quark-gluon plasma
is expected to be created in heavy ion collisions at the Relativistic Heavy Ion collider
(RHIC) and the Large Hadron Collider (LHC). The hot and dense quark matter might be
created in heavy ion collisions at FAIR in GSI and the Nuclotron-based Ion Collider Facility (NICA) at
JINR, while the cold and dense quark matter may exist in the inner
core of compact stars.

In the inner core of compact stars, the baryon density can reach or even be larger
than about 6 times
the normal nuclear matter density $n_0=0.16~\text{fm}^{-3}$, so there
might exist "exotic" matter like hyperons
\cite{Baldo:1999rq,Massot:2012pf,Whittenbury:2012rn}, meson condensations
\cite{Brown:1992ib,Glendenning:1998zx,Ramos:2000dq} and quark matter
(normal quark matter or strange quark matter \citep{Witten84,Farhi84},
and color superconductor \cite{Alford:2002rj,Shovkovy:2003ce}.).
Strange quark matter has been conjectured
to be the true ground state of QCD~\citep{Witten84,Farhi84}, and many efforts
have been put on investigating the conversion from neutron star to quark star
that consists of strange quark matter which is made purely by $u$, $d$, $s$
quarks and some leptons like electron and muon due to charge neutrality and $\beta$-equilibrium~\citep{Bom04,Sta07,Her11,Jaffe74}.
There are also hybrid star conjectures on a transition from nuclear phase to quark phase
at such high baryon density, and several authors have studied the phase transition in
hybrid star~
\citep{Col1975,Baym76,Freedman78,Drago96,Glen92,Prakash95,M¨¹ller96,Lattimer04,Steiner05}.

The equation of state (EoS) plays a central role in investigating properties of
strong interaction matter, which is one of fundamental issues in nuclear physics,
astrophysics and cosmology. On the one hand, the EoS generates unique
mass versus radius (M-R) relations for neutron stars and the ultra-dense remnants
of stellar evolution. On the other hand, the mass-radius relation of compact stars can
put strong constraints on the EoS for strong interaction matter at high
baryon density and low temperature.

Recently, two heaviest neutron stars have been measured with high accuracy.
One is the radio pulsar J1614-2230 \cite{Demorest:2010bx} with a mass of
$1.97\pm0.04 M_\odot$, and the other is J0348+0432 \cite{Antoniadis:2013pzd}
with mass $2.01\pm0.04 M_\odot$. Even heavier neutron
stars have been discussed in the literatures \cite{Lattimer:2006xb,Lattimer:2010uk}.
It is suggested that the established existence of two-solar-mass
neutron stars would prefer hard EoS based entirely on conventional
nuclear degrees of freedom, and many soft equations of state, including
hybrid stars containing significant "exotic" proportions of hyperons,
Bose condensates or quark matter would be ruled out.

However, it has been pointed out that the repulsive vector interaction in the
Nambu--Jona-Lasinio (NJL) model can produce a stiff EoS for dense
quark matter and thus can generate two-solar-mass compact star
\cite{Hell:2014xva,Menezes:2014aka}. The role of the vector interaction in
QCD vacuum and medium has been discussed in
many literatures \cite{Hell:2014xva,Menezes:2014aka ,Bernard:1995hm,Bernard:1987gw,Bernard:1987sg,Lutz:1992dv,Buballa:2003qv,Hanauske:2001nc,Huguet:2006cm,
Huguet:2007jc,Fukushima:2008wg,Abuki:2009ba,Zhang:2009mk,Bratovic:2012qs,Shao:2013toa,
Klahn:2013kga,Xu:2013sta,Steinheimer:2010sp,Steinheimer:2014kka}.
The introduction of the vector interaction within the NJL model is necessary
to describe vector mesons, and the coupling constant is determined by the vector spectra \cite{Bernard:1995hm,Lutz:1992dv}.
In order to describe vector bound states within the NJL model, the vector interaction
must be attractive in the space-like components, thus it is repulsive in the time-like
components, which is relevant for the number density in the mean field \cite{Buballa:2003qv}.
The QCD phase diagram and the critical end point is sensitive to the sign of vector
coupling constant as shown in \cite{Fukushima:2008wg,Abuki:2009ba,Zhang:2009mk,Bratovic:2012qs}.

Furthermore, the presence of external magnetic field can even harden the equation of
state of dense quark matter as shown in Ref. \cite{Menezes:2009uc}. In recent decades,
properties of hot and dense quark matter under strong magnetic fields have attracted
lots of interests, especially in recent several years much progress has been made.

Strong magnetic fields with the strength of $10^{18}\sim 10^{20} \textmd{G}$ (equivalent to $eB\sim (0.1-1.0~{\rm GeV})^2$) can be generated in the laboratory through non-central heavy ion collisions~\cite{Skokov:2009qp,Deng:2012pc} at the Relativistic Heavy Ion Collider (RHIC)
and the Large Hadron Collider (LHC). This offers a unique opportunity to study properties
of hot strong-interaction matter under strong magnetic field. The observation of charge azimuthal correlations at RHIC and LHC ~\cite{Abelev:2009ad,Abelev:2012pa} might indicate the anomalous
Chiral Magnetic Effect (CME) \cite{Kharzeev:2007tn,Kharzeev:2007jp,Fukushima:2008xe} with local $\mathcal{P}$- and $\mathcal{CP}$-violation. Conventional chiral symmetry breaking
and restoration under external magnetic fields has been investigated for many years. It has been recognized for more than 20 years that the chiral condensate increases with $B$, which is called
magnetic catalysis \cite{Klevansky:1989vi,Klimenko:1990rh,Gusynin:1995nb}, and naturally the
chiral symmetry should be restored at a higher $T_c$ with increasing magnetic field.
However, the lattice group \cite{Bali:2011qj,Bali:2012zg,Bali:2013esa} has demonstrated that the transition
temperature decreases as a function of the external magnetic field, i.e. inverse magnetic
catalysis around $T_c$, which is in contrast to the naive expectation and majority of
previous results. It was shown in \cite{Chao:2013qpa,Yu:2014sla} that the chirality imbalance
can explain the inverse magnetic catalysis. It was suggested that in the presence of
external magnetic field, there will be vector condensation in the QCD vacuum \cite{Chernodub:2010qx,Chernodub:2011mc}, and
the vector condensation was confirmed in Refs. \cite{Frasca:2013kka,Liu:2014uwa}.
The vector condensation in the neutron star has been discussed in Ref.\cite{Mallick:2014faa}.

Moreover, considerable efforts have also been directed to the study of the effects of
intense magnetic fields on various astrophysical phenomena.
The presence of strong magnetic fields at the surface of conventional compact stars
or neutron stars is $10^9\sim10^{15} \textmd{G}$ ~\citep{Lyne05,Woltjer64,Miharaet90,Chanmugam92,Thom96,Ibrahim02,Dun92},
which is thousand times stronger than ordinary neutron stars.
These strongly magnetized objects are called magnetars~\citep{Dun92}.
By using scalar virial theorem
based on Newtonian gravity~\citep{Lai91}, it is predicted that the magnetic field in
the inner core of neutron stars could reach as high as $10^{18}\sim 10^{20}\textmd{G}$.
Under such tremendous magnetic fields, the $\mathcal{O}(3)$ rotational symmetry will
break and the pressure anisotropy of the system must be considered \citep{Ferrer:20102013,Isayev11,Isayev12,Isayev13}. In order to mimic the
spatial distribution of the magnetic field strength in magnetars, people
have introduced a density-dependent magnetic field profile \citep{Ban97,Ban98}.

Many efforts have been taken on investigating the existence of quark core in neutron stars
under strong magnetic field \cite{Menezes:2008qt,Menezes:2009uc,Casali:2013jka,Strickland:2012vu,Sinha:2013dfa,Denke:2013gha,
Dexheimer:2011pz,Chu:2014foa}.
In this work, we will use SU(3) Nambu-Jona-Lasinio(NJL) model with vector interaction
to investigate the magnetar, and we will also consider the pressure contribution
from polarized gluons under magnetic field.

It is known that the NJL model only considers the quark contribution to the pressure,
therefore, the pressure from NJL model is smaller than that from lattice calculation at high
temperature and zero chemical potential \cite{Zhuang:1994dw,Boyd:1996bx}. In order to consider
the gluon contribution to the pressure, the Polyakov-loop potential was
firstly introduced in the framework of NJL model in \cite{Fukushima:2003fw}
and then in \cite{Ratti:2005jh}, where the Polyakov-loop potential is temperature
dependent. The PNJL model can fit the equation of state at high temperature very well with
lattice QCD results. The extension of the Polyakov-loop potential to finite chemical
potential is not trivial. In order to use the PNJL model to describe neutron stars, the
Polyakov potential is modified to have chemical potential dependent in Refs. \cite{Dexheimer:2009hi,Dexheimer:2009va,Blaschke:2010vj,Shao:2011nu,Lourenco:2012yv}.
However,  there is still no extension of Polyakov-loop potential under magnetic field.
Therefore, we cannot use PNJL model for our purpose in this work to investigate quark matter
at high baryon density under external magnetic field, thus we have to find other way to include
the gluon contribution to the pressure at finite chemical potential and at magnetic fields.
In this work, we will give an ansatz on the potential from magnetized gluons hinted from hard
thermal/dense loop results.

The paper is organized as follows. In Sec. II, we give a general description of the SU(3)
NJL model with vector interaction under magnetic field with $\beta$ equilibrium, we
make an ansatz of thermodynamical potential from magnetized gluons, and derive
the equation of state of the strange quark matter with $\beta$-equilibrium. Our numerical
results are shown in Sec. III and the conclusion and discussion is given in Sec. IV.

\section{Three-flavor NJL model with vector interaction under magnetic field}
\label{Sec-NJLModel}

We study properties of three-flavor system with external strong magnetic fields $A_{\mu}^{ext}$
under $\beta$ equilibrium condition, which is described by the Lagrangian density
\begin{equation}
\mathcal{L} = \mathcal{L}_{q}+\mathcal{L}_e-\frac{1}{4}F_{\mu\nu}F^{\mu\nu},
\label{L}
\end{equation}
where $\mathcal{L}_q$, $\mathcal{L}_e$ are the Lagrangian densities for quarks
and electrons, respectively. $F_{\mu\nu}=\partial_{\mu} A_{\nu}^{ext}-\partial_{\nu} A_{\mu}^{ext}$ is the strength tensor for external electromagnetic field. The magnetic field $B$ is a static magnetic field along z direction, and $A_\mu^{ext}=\delta_{\mu2}x_1 B$. In this work,
we do not consider contributions from the anomalous magnetic moments \citep{Duncan00}.

The electron Lagrangian density is given as
\begin{equation}
\mathcal{L}_e=\bar{e}[(i\partial_{\mu}-eA^{\mu}_{ext})\gamma^{\mu}]e.
\end{equation}
The Lagrangian density for quarks is described by the gauged $N_f=3$ NJL model
with vector interaction \cite{Bernard:1987gw,Bernard:1987sg}
\begin{equation}
\mathcal{L}_{q}=\bar{\psi}_f[\gamma_{\mu}(i\partial^ \mu-q_f A^{\mu}_{ext}) - \hat{m}_c]\psi _f +\mathcal{L}_ {4} +\mathcal {L}_{6},
\end{equation}
where $\mathcal{L}_ {4}$ indicates four-fermion interaction compatible with QCD symmetries $SU(3)_{color}\otimes SU(3)_L\otimes SU(3)_R$,
and $\mathcal {L}_{6}$ is the six-point interaction which is required to break the axial $U(1)_A$ symmetry. $\psi=(u,d,s)^T$ represents a quark field with three flavors,
$\hat{m}_c={\rm diag}(m_u,m_d,m_s)$ is the current quark mass matrix, and $q_f$ the quark
electric charge.
The four-fermion interaction includes scalar, pseudoscalar, vector, axial-vector channels
and takes the form of
\begin{equation}
\mathcal{L}_ {4}=\mathcal{L}_{S}+\mathcal{L}_{V}+\mathcal{L}_{I,V}.
\end{equation}
The scalar part takes the form of
\begin{equation}
\mathcal{L}_{S} = G_S\sum\limits _ {a=0}^8 [(\bar{\psi _f}\lambda_a \psi_f)^2 +(\bar{\psi_f}i\gamma_5\lambda_a\psi_f)^2],
\end{equation}
and the vector part is given as
\begin{equation}
\mathcal{L}_V=-G_V{\sum\limits_{a=0}^8[(\bar{\psi}\gamma^{\mu}\lambda^a\psi)^2
+(\bar{\psi}i\gamma^\mu\gamma_5\lambda^a\psi)^2]},
\end{equation}
where $G_S$ and $G_V$ are the coupling constants in the scalar and vector channels,
respectively. $\lambda_a (a=1,\cdots,8)$ $\lambda_a $ are the Gell-Mann matrices, and
the generators of the SU(3) flavor groups,
and $\lambda_0=\sqrt{2/3}I$ with $I$ the $3\times 3$ unit matrix.
In order to describe the nonet of scalars, pseudo-scalars, vectors and axial-vectors,
a convenient representation is obtained by changing from $\{\lambda_0,\lambda_1,
\cdots,\lambda_8\}$ to the set $\{\lambda_0,\lambda_1^{\pm},\lambda_3,\lambda_4^{\pm},\lambda_6^{\pm},\lambda_8\}$ with
\begin{eqnarray}
\lambda_1^{\pm}&=&\sqrt{\frac{1}{2}}(\lambda_1 \pm i \lambda_2), \nonumber \\
\lambda_4^{\pm}&=&\sqrt{\frac{1}{2}}(\lambda_4 \pm i \lambda_5), \nonumber \\ \lambda_6^{\pm}&=&\sqrt{\frac{1}{2}}(\lambda_6 \pm i \lambda_7). \nonumber
\end{eqnarray}

Hadrons in the $u,d$ sector exhibit $SU(2)_I$ isospin symmetry. Up and down quarks have isospin
$I = 1/2$, and isospin 3-components ($I_3$) of $1/2$ and $-1/2$, respectively. All other quarks
have $I = 0$. For scalars, the coupling constant in the scalar-isoscalar ($\sigma$) and pseudoscalar-isovector
($\pi$) interactions have to be equal, which is constrained by chiral symmetry.
However, the coupling constants for the vector-isoscalar ($\omega$) and vector-isovector ($\rho$) interaction
can be separately invariant, thus can be chosen independently. The ratio of the coupling constants of the
vector-isosinglet channel $\omega$ and vector-isovector channel $\rho$ to nucleons is empirically given
by $g_{\omega QQ}/g_{\rho QQ}\simeq 3$ in the chirally broken phase, and $g_{\omega QQ}/g_{\rho QQ}= 1$
in the chiral symmetric phase \cite{Abuki:2009ba,Duncan00,Sasaki:2006ws}. To distinguish the isoscalar
and isovector for vectors, we introduce an extra term for vector-isovector channel with the form of
\begin{equation}
\mathcal{L}_{IV}=-G_{IV}[(\bar{\psi}\gamma^{\mu}\vec\tau\psi)^2
        +(\bar{\psi}\gamma_5\gamma^{\mu}\vec\tau\psi)^2].
\end{equation}
The coupling constant for vector-isoscalar $G_V^{\omega}=G_V$, and the coupling
constant for vector-isovector $G_V^{\rho}=G_V+G_{IV}$. In this work, we will investigate
the role of vector-isovector interaction on the equation of state, therefore in our
numerical calculations, we choose $G_{IV}$ as a free parameter.

The six-fermion interaction $\mathcal{L}_{6}$, i.e. the 't Hooft term takes the form of
\begin{equation}
 \mathcal{L}_{det} = -K\{{\det}_f[\bar{\psi_f} (1+\gamma_5)\psi_f]+{\det}_f{[\bar{\psi_f }(1-\gamma_5)\psi_f]}\},
\end{equation}
which is to break the $U(1)_A$ symmetry.

\subsection{The pressure from quark contribution}

The equation of state is the most important aspect for physicists to acquire
the properties of quark matter. In order to get the pressure and energy density
of quark matter, we should firstly derive the thermodynamical potential $\Omega_f$.
In procedure, we can calculate the thermodynamical quantities by using finite
temperature field theory. In the mean-field approximation, the lagrangian
density for quark part is
\begin{eqnarray}
\mathcal{L}_{M}&=&\bar{\psi}_f[\gamma_{\mu}(i\partial^ \mu-q_f A^{\mu}_{ext})-\hat{M} -2G_{IV}\gamma_0\tau_{3f}n_f]\psi_f \notag \\
&-&2G_S(\sigma_u^2+\sigma_d^2+\sigma_s^2)+4K\sigma_u\sigma_d\sigma_s
    -4G_V\gamma_0\hat{n} \notag \\
&+&2G_V(n_u^2+n_d^2+n_s^2)+G_{IV}(n_u-n_d)^2, \nonumber \\
\end{eqnarray}\\
where
$
\hat{n}=\left( \begin{array}{ccc}
n_u&0&0\\0&n_d&0\\0&0&n_s
\end{array} \right)
$ and $\hat{M}=\left( \begin{array}{ccc}
M_u&0&0\\0&M_d&0\\0&0&M_s
\end{array} \right). $\\ \\

The quark mass is determined by the gap equation of
\begin{equation}
M_i=m_i-4G_S\sigma_i+2K\sigma_{j}\sigma_{k},
\end{equation}
with $(i,j,k)$ being any permutation of $(u,d,s)$ and the chiral condensate is given as
\begin{equation}
\sigma_f=\langle\bar{\psi_f}\psi_f\rangle=-i\int \frac{d^4p}{(2\pi)^4}\text{tr}\frac{1}{(p\!\!\!/-M_f+i\epsilon)}.
\end{equation}

After introducing Landau quantization and several steps of finite temperature field theory calculations, we can get the thermodynamical potential $\Omega_q$ of quark matter under
magnetic fields, and the pressure density $p_q=-\Omega_q$ and takes the form of
\begin{eqnarray}
p_q&=&- 2G_S(\sigma_u^2+\sigma_d^2+\sigma_s^2)+4K\sigma_u\sigma_d\sigma_s\notag \\&+&2G_V(n_u^2+n_d^2+n_s^2) + G_{IV}(n_u-n_d)^2 \nonumber \\
&+& (\Omega_{ln}^u+\Omega_{ln}^d+\Omega_{ln}^s )
\label{pq}
\end{eqnarray}
with the logarithmic contribution
\begin{equation}
\Omega_{ln}^f=-i\int\frac{d^4p}{(2\pi)^4}
\text{tr}\ln{\{\frac{1}{T}[p\!\!\!/-\hat{M_f}+\gamma_0\tilde{\mu}_f]}\},
\end{equation}
here,
\begin{equation}
\tilde{\mu}_f=\mu_f-4G_Vn_f -2G_{IV}\tau_{3f}(n_u-n_d),
\end{equation}
where $\mu_f$ is the chemical potential for each flavor of quarks and $\tau_{3f}$ is the isospin quantum number for quarks: $\tau_{3u}=1$, $\tau_{3d}=-1$ and $\tau_{3s}=0$.

Following Ref.\cite{Menezes:2009uc}, one can get the condensates and pressure
for quarks. The logarithmic contribution to the thermodynamical potential is given by
\begin{eqnarray}
\Omega_{ln}^f&=&\Omega_{ln}^{f,vac}+\Omega_{ln}^{f,mag}+\Omega_{ln}^{f,med}.
\end{eqnarray}
The first term is the vacuum contribution
\begin{eqnarray}
\Omega_{ln}^{f,vac}=-\frac{N_c }{8\pi^2}\Big\{M_f^4\ln{\Big[\frac{\Lambda+\epsilon_{\Lambda}}{M_f}\Big]
-\epsilon_{\Lambda}\Lambda(\Lambda^2+\epsilon_{\Lambda}^2)}\Big\}
\end{eqnarray}
with $\epsilon_{\Lambda}^2=\Lambda^2+M_f^2$ and $\Lambda$ the noncovariant cutoff.
The magnetic field contribution takes the form of
\begin{eqnarray}
\Omega_{ln}^{f,mag}&=& \frac{N_c}{2\pi^2} (|{q_f}|B)^2\Big[\frac{x_f^2}{4}+\zeta'(-1,x_f)
  \nonumber \\
& &-\frac{1}{2}(x_f^2-x_f)\ln{(x_f)}\Big],
 \end{eqnarray}
where $\zeta(z,x)$ is the Riemann-Hurwitz function and
\begin{equation}
\zeta'(-1,x_f)=d\zeta(z,x)/dz\vert _{z=-1}
\end{equation}
with $x_f=\frac{M_f^2}{2| {q_f}|B} $.
The medium contribution reads
\begin{eqnarray}
\Omega_{ln}^{f,med}&=& \sum\limits_{k=0}^{k_{fmax}}\alpha_k\frac{(|{q_f}|BN_c)}{4\pi^2}
\Big\{\tilde{\mu}_f\sqrt{\tilde{\mu}_f^2-s_f(k,B)^2}
\notag \\&-&s_f(k,B)^2\ln{\Big[\frac{\tilde{\mu}_f+\sqrt{\tilde{\mu}_f^2-s_f(k,B)^2}}{s_f(k,B)}\Big]}\Big\},\notag\\
\end{eqnarray}
where \begin{equation}
s_f(k,B)=\sqrt{M^2+2|{q_f}|B k},
\end{equation}
and
\begin{equation}
k_{fmax}=\frac{\tilde{\mu}_f^2-M^2}{2|{q_f}|B}=\frac{p^2_{f,F}}{2|{q_f}|B},
\end{equation}
is the upper Landau level with $\alpha _k= 2-\delta_{k0}$.

Then we can also give the condensates for each flavor of quarks:
\begin{eqnarray}
\sigma_f=\sigma_f^{vac}+\sigma_f^{mag}+\sigma_f^{med}
\end{eqnarray}
with
\begin{eqnarray}
\sigma_f^{vac}&=&-\frac{M_fN_c}{2\pi^2}\Big\{\Lambda\sqrt{\Lambda^2+M_f^2}\notag \\&-&\frac{M_f^2}{2}\ln{\Big[\frac{(\Lambda+\sqrt{\Lambda^2+M_f^2})^2}{(M_f^2)}\Big]}\Big\},\\
\sigma_f^{mag}&=&-\frac{M_fN_c}{2\pi^2}(|{q_f}|B)\Big\{\ln[\Gamma(x_f)]\notag \\ &-&\frac{1}{2}\ln(2\pi)+\frac{\ln (x_f)}{2}-x_f\ln (x_f)\Big\}, \\
\sigma_f^{med}&=&\sum\limits_{k=0}^{k_{fmax}}\alpha_k\frac{M_f|{q_f}|BN_c}{\pi^2}\notag \\
& & \Big\{\ln{\Big[\frac{\tilde{\mu}_f+\sqrt{\tilde{\mu}_f^2-s_f(k,B)^2}}{s_f(k,B)}\Big]}\Big\}.
 \end{eqnarray}

\subsection{The pressure from leptons}

For SQM, we assume it is neutrino-free and composed of $u$, $d$, $s$ quarks
and $e^-$ in beta-equilibrium with electric charge neutrality. The weak
beta-equilibrium condition can then be expressed as
\begin{eqnarray}
&&\mu_u+\mu_e=\mu_d=\mu_s,
\end{eqnarray}
where $\mu_i$ ($i=u$, $d$, $s$ and $e^-$) is the chemical potential of the
particles in SQM. Furthermore, the electric charge neutrality condition can
be written as
\begin{eqnarray}
\frac{2}{3}n_u=\frac{1}{3}n_d+\frac{1}{3}n_s+n_e.
\end{eqnarray}
Where
\begin{equation}
n_f=\sum\limits_{k=0}^{k_f,max}\alpha_k\frac{|{q_f}|BN_c}{2\pi^2}k_F,_f
\end{equation}
is the number density for each flavor of quarks with $k_{F,f}=\sqrt{\tilde\mu_f^2-s_f(k,B)^2}$,
and
\begin{equation}
n_l=\sum\limits_{k=0}^{k_l,max}\alpha_k\frac{|{q_l }|B}{2\pi^2}k_{F,_l}
\end{equation}
the number density of electrons.

We can also write the leptonic contribution to the pressure density, which
takes the form of:
\begin{eqnarray}
p_l&=& \sum\limits_{k=0}^{k_{lmax}}\alpha_k\frac{(|{q_l}|BN_c)}
{4\pi^2}\Big\{\mu_l\sqrt{\mu_l^2-s_l(k,B)^2}\notag \\&-&s_l(k,B)^2\ln{\Big[\frac{\mu_l+\sqrt{\mu_l^2-s_l(k,B)^2}}{s_l(k,B)}\Big]}\Big\}.
\end{eqnarray}

\subsection{The pressure from magnetized gluon potential}

It is well known that the NJL model only considers the quark contribution to the
pressure, which is smaller than that from lattice calculation at high
temperature and zero chemical potential \cite{Zhuang:1994dw,Boyd:1996bx}.
In the presence of strong magnetic field, not only quarks are polarized along
the direction of $B$, but also gluons will be polarized via the quark loop.
There are still few calculations of pressure contributed from magnetized gluon
degrees of freedom at zero temperature and finite chemical potential \cite{Bali:2013esa,Levkova:2013qda}.

In order to consider the gluon contribution to the pressure,
on the one hand, the Polyakov-loop potential was introduced in the framework of NJL
model in \cite{Fukushima:2003fw,Ratti:2005jh}, where the
Polyakov-loop potential is temperature dependent. The PNJL model can fit the equation of
state at high temperature very well with lattice QCD results. However, to extend the
Polyakov-loop potential to finite chemical potential and strong magnetic field
is nontrivial, though a modified version of the Polyakov potential at finite chemical
potential has been proposed in Refs. \cite{Dexheimer:2009hi,Dexheimer:2009va,Blaschke:2010vj,Shao:2011nu,Lourenco:2012yv}.
For our purpose in this work to describe the magnetar, we cannot use the PNJL model,
but we can estimate the gluon contribution to the pressure at finite chemical potential
and with magnetic fields from hard-thermal/dense-loop results. At high
temperature and zero chemical potential with zero magnetic field, the gluon contribution
to the pressure in the PNJL model should merge with hard-thermal-loop result.

Much efforts have been put on the perturbation theory (PT) of
hard-thermal-loop (HTL) or hard-dense-loop (HDL) calculation
on the equation of state of strong interaction matter at high temperature and density
\cite{Freedman:1976ub,Baluni:1977ms,Blaizot:2000fc,Fraga:2001id,Fraga:2001xc,Andersen:2002jz,
Vuorinen:2003fs,Fraga:2004gz,Ipp:2006ij,Kurkela:2009gj,Kurkela:2014vha,Haque:2014rua,
Mogliacci:2014pxa}. At high temperature, recent progress up to three-loop HTL calculations \cite{Haque:2014rua} shows that the pressure density, energy density and other thermodynamical
properties are in good agreement with available lattice data for temperatures above approximately
$300 {\rm MeV}$. The EoS of cold quark matter is accessible through
perturbative QCD at high densities, and has been determined to order
$\alpha_s^2$ in the strong coupling constant \citep{Kurkela:2009gj}.

The pressure density for ideal gas of quarks and gluons has the form of
\begin{eqnarray}
p^{SB}&=& p_q^{SB}+p_g^{SB}, \\
p_q^{SB} & = & N_cN_f \Big(\frac{7\pi^2T^4}{180}+\frac{\mu^2T^2}{6}+\frac{\mu^4}{12\pi^2}\Big), \\
p_g^{SB} &=& (N_c^2-1)\frac{\pi^2T^4}{90}.
\end{eqnarray}
It is noticed that the pressure density for the ideal gas of gluons is only temperature
dependent and with no chemical potential dependent.
Switching on coupling between quarks and gluons, the gluons will get screening mass
\begin{equation}
m_g^2=\frac{1}{6}\Big[(N_c+\frac{1}{2}N_f)T^2+\frac{3}{2\pi^2}\sum_f\mu_f^2\Big]g_{eff}^2,
\end{equation}
and the perturbation theory at one-loop gives the pressure density of gluons as
\begin{equation}
p_g^{PT}=p_g^{SB}-N_g\frac{g_{eff}^2}{32}\Big[\frac{5}{9}T^4+\frac{2}
{\pi^2}\mu^2T^2+\frac{1}{\pi^4}\mu^4\Big],
\label{pgPT}
\end{equation}
the gluon pressure becomes chemical potential dependent via quark loop, though the
coefficient in front of $\mu^4$ is only 1/54 of that of $T^4$.

Under the strong magnetic field, a massive resonance $\sim g\sqrt{eB}$ is excited in the
longitudinal $(1+1)$ component of the gluon propagator under strong magnetic fields
 \cite{miransky2002magnetic}. The corresponding \textit{Debye mass} of
the longitudinal gluon fields $A_{\shortparallel}$ has the screening mass
\begin{equation}
m_g^2(eB)=  \sum_f |q_f| \frac{g_{eff}^2}{4 \pi^2}|eB|,
\end{equation}
at zero temperature and density. One can guess that at nonzero temperature and density,
the longitudinal gluon fields take the screening mass of
\begin{equation}
m_g^2(T,\mu,eB)=g^2(a T^2+b \mu^2 + c eB)
\end{equation}
where $a,b,c$ are constants. Taking into account the transverse gluons,
the pressure density from magnetized gluons can be estimated as
\begin{equation}
p_g(T,\mu;eB)=a_0\mu^2eB+b_0\mu^4+c_0T^2eB+d_0\mu^2T^2+e_0T^4,
\end{equation}
with $a_0,b_0,c_0,d_0,e_0$ being free parameters. At zero temperature,
\begin{equation}
p_g(T=0,\mu;eB)=a_0\mu^2eB+b_0\mu^4.
\end{equation}
Furthermore, for the screening mass of gluons, considering the coefficients in
front of $eB$ is almost the same magnitude as that of $\mu^2$, and we neglect the
anisotropy pressure density caused by the magnetic field with magnitude $eB <10^{19} G$, for simplicity, in this work, we use the following ansatz of the pressure density
of magnetized gluons
\begin{equation}
p_g(T=0,\mu;eB)=a_0(\mu^2eB+\mu^4).
\label{pgnonPT}
\end{equation}
for our numerical calculations. If we directly extend the perturbative gluon
pressure Eq.(\ref{pgPT}) to nonperturbative region, the gluon pressure density
at zero temperature and finite density should be negative. However, in the moderate
baryon density region, nonperturbative feature of gluons should still play an
important role as shown in \cite{Megias:2005ve,Fukun}. The system in the moderate
baryon density can be regarded as compositions of quasi-quarks described by NJL model
and quasi-gluons. To compensate the quasi-quark contribution to the pressure in
the NJL model, the quasi-gluon contribution to the pressure should be also positive.
In our numerical calculation, we will take $a_0>0$ for physical case, but
we will also take $a_0<0$ for reference.

\subsection{Total pressure of SQM with $\beta$-equilibrium under magnetic field}

Under strong magnetic fields, the $\mathcal{O}(3)$ rotational symmetry
in SQM is broken and the pressure for SQM might become anisotropic, i.e., the
longitudinal pressure $P_{||}$ which is parallel to the magnetic field orientation
is different from the transverse pressure $P_{\perp}$ which is perpendicular to
the orientation of magnetic field. The analytic forms for longitudinal and
transverse pressure densities of the system are given
by \cite{Ferrer:20102013}
\begin{equation}
p_{||}=p-\frac{1}{2}B^2,
\label{pP}
\end{equation}
\begin{equation}
p_{\perp}=p+\frac{1}{2}B^2-{M}B,
\label{pT}
\end{equation}
where we have defined
\begin{equation}
p=p_q+p_l+p_g-p_0,
\end{equation}
with $p_0=-\Omega_0=-\Omega(T=0,\mu=0,B=0)$ the vacuum pressure density,
which ensures $p=0$ in the vacuum.
$M$ is the system magnetization, and takes the form of
\begin{align}
M=-\partial{\Omega}/\partial{B}=\sum\limits_{i=u,d,s,l,g} M_i.
\end{align}
The energy density for SQM at zero temperature is given by
\begin{equation}
\epsilon=-p+\sum\limits_{i=u,d,s,l}\mu_in_i+\frac{1}{2}B^2.
\end{equation}

One can find that the longitudinal pressure density $p_{||}$ satisfies the Hugenholtz-Van Hove(HVH)
theorem \cite{HVH,Chu:2014foa}, while the transverse pressure density $p_{\perp}$ does not because of the extra contributions from the magnetic field. We can see that the magnetic energy density term $B^2/2$ contributes oppositely to the longitudinal and transverse pressures under magnetic fields,
which will lead to a tremendous pressure anisotropy when the magnetic field is very strong.

\section{Numerical results and Conclusions}

For our numerical calculations, following \cite{Menezes:2009uc}, the set of
parameters we used is: $\Lambda=631.4$MeV, $m_u=m_d=5.5$MeV, $m_s=135.7$MeV,
$G\Lambda^2=1.835$ and $K\Lambda^5=9.29$.

\subsection{The effect of vector-isovector interaction and vector-isoscalar interaction
under magnetic field}

Firstly, we analyze the effect of vector-isovector and vector-isoscalar interaction
on the equation of state of strange quark matter Eq.~(\ref{pq}) under zero magnetic
field $B=0$. For zero magnetic field, there is no anisotropy in the system, the
longitudinal pressure density is equal to the transverse pressure density, i.e.,
$p_{||}=p_{\perp}=p(B=0)$.

\begin{figure}[tbp]
\includegraphics[scale=0.35]{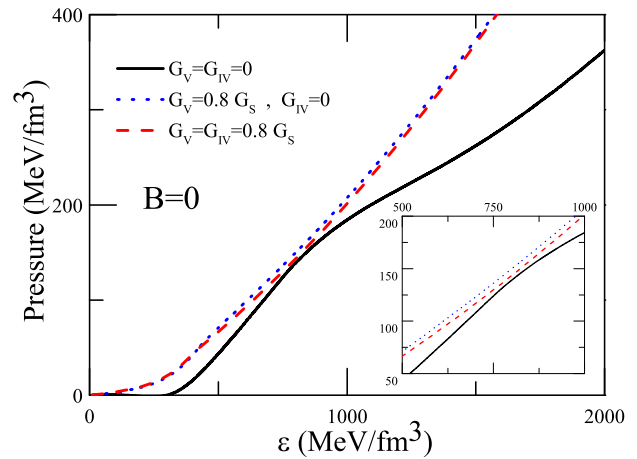}
\caption{(Color online) The pressure density of SQM as a function of energy density under zero
magnetic field with three cases: $G_V=G_{IV}=0$, $G_V=0.8G_S,G_{IV}=0$ and $G_V=G_{IV}=0.8G_S$.}
\label{Fig:GIV}
\end{figure}

In Fig.\ref{Fig:GIV}, we show the pressure density of SQM as a function
of energy density for three cases: 1) $G_V=G_{IV}=0$, 2) $G_V=0.8G_S,G_{IV}=0$, and 3) $G_V=G_{IV}=0.8G_S$.
Comparing the cases of 1) $G_V=G_{IV}=0$ and 2) $G_V=0.8G_S,G_{IV}=0$,
one can find that the repulsive vector-isoscalar interaction $G_V$ gives a stiffer
equation of state. However, comparing the cases of 2) $G_V=0.8G_S,G_{IV}=0$ and
3) $G_V=G_{IV}=0.8G_S$, it is observed that the equation of states for these two cases are
almost the same, which indicates that the vector-isovector interaction does not affect
the equation of state too much in SQM. One can find from Eq.~(\ref{pq}) that the contribution
from the vector-isovector interaction is mainly dependent on the u-d quark isospin
asymmetry($n_u-n_d$) and the coupling constant $G_{IV}$. Since the isospin asymmetry in SQM is small, the vector-isovector interaction is very tiny with the parameter set $G_V=G_{IV}=0.8G_S$.

It can be understood like that, the repulsive vector-isoscalar interaction shifts
the chemical potential to a larger value, which makes the equation of state stiffer.
However, the interaction in the vector-isovector channel shifts the isospin
chemical potential, and this effect is negligible for the equation of state under $\beta$-equilibrium. Therefore, in the following numerical calculations, we simply
take $G_{IV}=0$.

\begin{figure}[tbp]
\includegraphics[scale=0.42]{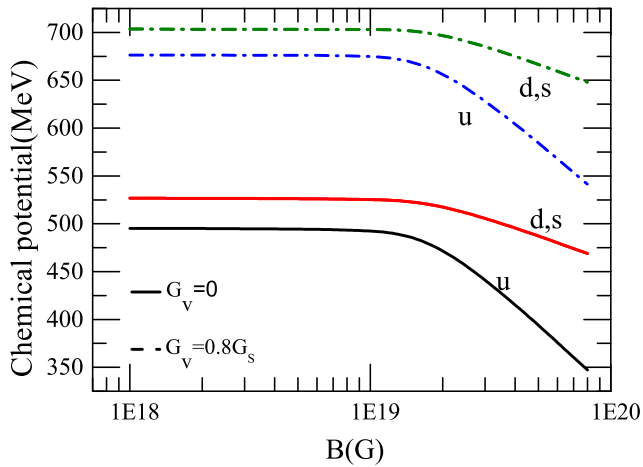}
\caption{(Color online)Chemical potentials for $u,d,s$ quarks as functions of magnetic
field with $G_V=0$ and $G_V=0.8G_S$ at baryon density $n_b=10 n_0$ in SQM. }
\label{chemical}
\end{figure}

Fig. \ref{chemical} shows the chemical potentials for u, d and s quarks as functions of
the magnetic field $B$ with $G_V=0$ and $G_V=0.8G_S$ at fixed baryon number density
$n_b=10 n_0$ in SQM. The chemical potential for each
flavor with $G_V=0.8G_S$ is enhanced magnificently comparing the case of $G_V=0$,
which implies that a stiffer EoS can be generated once considering large vector-isoscalar
coupling constant. One can also observe that, for both cases with $G_V=0$ and $G_V=0.8G_S$,
the chemical potential for quarks keeps a constant below the magnitude of $10^{19}{\rm G}$,
and decreases with the constant magnetic field above that.

\begin{figure}[tbp]
\includegraphics[scale=0.44]{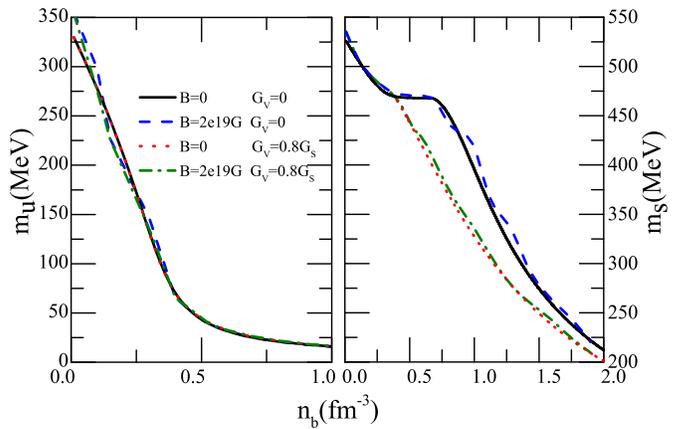}
\caption{(Color online) Constituent mass for u quark (left figure) and s quark (right figure)
as functions of baryon number density in charge neutral SQM for $B=0~\text{and}~2\times10^{19}G$
with $G_V=0~\text{and}~0.8G_s$, respectively.}
\label{mu}
\end{figure}

Fig.~\ref{mu} shows the constituent mass of u quark and s quark as functions of
baryon density in charge neutral SQM for $B=0$ and $B=2\times10^{19} G$ with $G_V=0$
and $G_V=0.8G_S$, respectively. In the case of zero magnetic field $B=0$ and $G_V=0$,
the constituent quark mass for u quark decreases from the vacuum mass almost linearly
to $50 {\rm MeV}$ in the region of baryon number density below
$n_b\simeq 0.35 {\rm fm}^{-3}\simeq 2 n_0$, and then slowly decreases with baryon number
density. The constituent quark mass for s quark also drops almost linearly from its vacuum
mass to around $475 {\rm MeV}$ in the region of baryon number density below
$n_b\simeq 0.35 {\rm fm}^{-3}\simeq 2 n_0$, and almost keeps as a constant
in the region of $ 0.35 <n_b < 0.7 {\rm fm}^{-3}$ ( $ 2 <n_b/n_0 < 4 $),
and then decreases quickly with baryon number density. This behavior is similar to
Fig. 3.9 in \cite{Buballa:2003qv}. When the magnetic field is turned on, it is found
that under the magnitude of $B=2\times10^{19} G$, only a tiny magnetic catalysis effect
can be observed for u quark and s quark in the low baryon density region, and at high
baryon density region, the magnetic field with magnitude of $B=2\times10^{19} G$ almost
has no effect on constituent quark mass at fixed baryon density. However, it is noticed
that the repulsive vector-isoscalar interaction can smooth away the saturation region
and make the strange quark mass decreases linearly with the baryon number density.

\begin{figure}[tbp]
\includegraphics[scale=0.42]{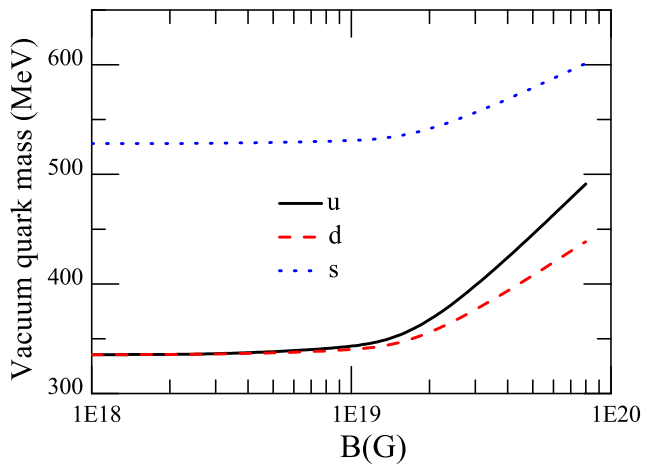}
\caption{(Color online) Vacuum constituent masses for u, d and s quarks as functions of
the magnetic field in SQM.}
\label{Vacuumquark}
\end{figure}

As shown in Fig. \ref{mu}, the magnetic field in the magnitude of
$B=2\times10^{19} G$ does not essentially affect the constituent quark mass.
In Fig. \ref{Vacuumquark}, we investigate the vacuum constituent mass of
$u,d~\text{and}~s$ quarks as functions of magnetic field in SQM, and one
can find that the masses of three different flavors of quarks
do not change so much when the magnetic field is smaller than $10^{19} G$,
while the $u,d$ quark masses increase drastically when the magnetic field is
bigger than $3\times 10^{19} G$, which indicates the magnetic catalysis phenomenon.
It should also be noticed that the masses of $d$ quark and $s$ quark increase more
slowly with the magnetic field compared to the u quark mass case.

\subsection{The equation of state, sound velocity and magnetar mass}

As it is accepted, the magnetic field strength in the inner core region of compact stars
could be much larger than the magnetic field at the surface, then a
density-dependent magnetic field distribution inside the compact star is usually
used to describe this behavior. We use the following
popular parametrization for the density-dependent magnetic field
profile in QSs as Refs.\citep{Ban98, Menezes09, Menezes209, Ryu10, Ryu12}.
\begin{eqnarray}
B=B_{surf}+B_0[1-\exp{(-\beta_0(n_b/n_0)^\gamma)}],
\end{eqnarray}
where $B_{surf}$ is the magnetic field strength at the surface of
compact stars and its value is fixed at $B_{surf} = 10^{15} G$ in this work,
$n_0 = 0.16 fm^{-3}$ is the normal nuclear matter density, $B_0$ is the constant
magnetic field, which is a parameter with dimension of $B$, and $\beta_0 $ and
$\gamma$ are two dimensionless parameters that control how exactly the magnetic
field strength decays from the center to the surface. In order to reproduce
the magnetic field which is weak below the nuclear saturation point while
getting stronger at higher density, we take $B_0=4\times 10^{18}G,~\beta=0.003$ and $\gamma=3$
as the set of parameters in the following calculations, and this magnetic field distribution is already proved to be a gentle magnetic field distribution for SQM inside QSs, which can lead
to a small pressure anisotropy and small maximum mass splitting for QSs in a
density-dependent quark model\citep{Chu:2014foa}.

\begin{figure}[tbp]
\includegraphics[scale=0.42]{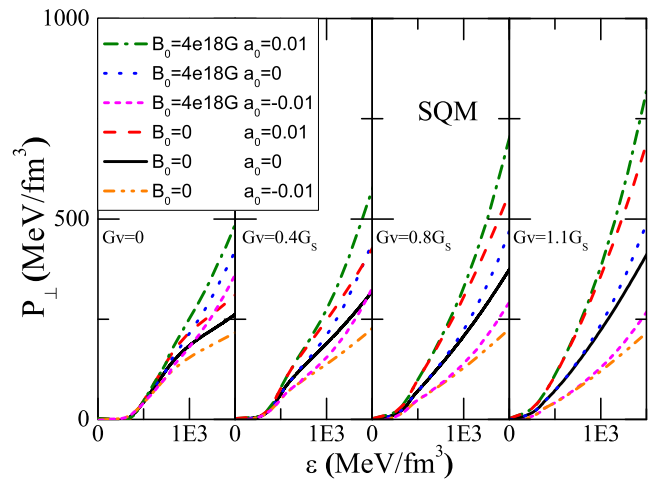}
\caption{(Color online) Transverse pressure density as a function of energy density for
SQM in cases of $B_0=0$ and $B_0=4\times 10^{18}G$
with $a_0=0,0.01,-0.01$ and $G_V=0$, $G_V=0.4G_S$, $G_V=0.8G_S$ and $G_V=1.1 G_S$,
respectively. }
\label{PE}
\end{figure}

\begin{figure}[tbp]
\includegraphics[scale=0.43]{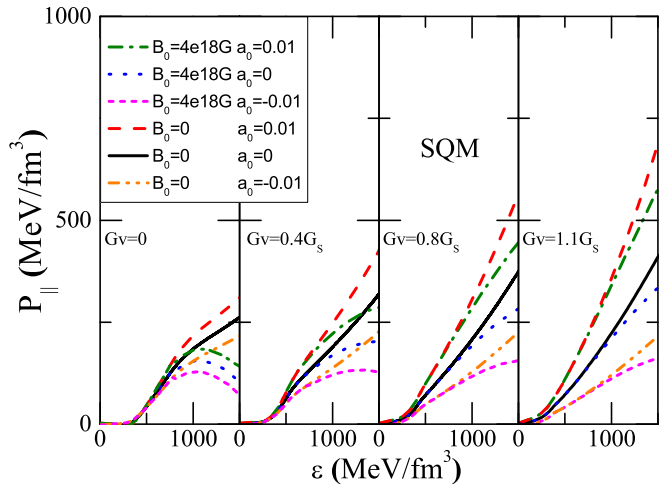}
\caption{(Color online)Longitudinal pressure density as a function of energy density SQM
in cases of $B_0=0$ and $B_0=4\times 10^{18}G$
with $a_0=0,0.01,-0.01$ and $G_V=0$, $G_V=0.4G_S$, $G_V=0.8G_S$ and $G_V=1.1 G_S$,
respectively. }
\label{PLE}
\end{figure}

The anisotropic pressure densities in Eq.(\ref{pP}) and Eq.(\ref{pT})
are calculated in cases of $B_0=0$ and $B_0=4\times 10^{18}G$ with $a_0=-0.01,0,0.01$ and
$G_V=0$, $G_V=0.4G_S$, $G_V=0.8G_S$ and $G_V=1.1 G_S$, respectively. We calculate the
transverse pressure density as a function of energy density for
SQM in Fig.~\ref{PE} while the longitudinal pressure case in Fig.~\ref{PLE}.


One can see from Fig.~\ref{PE} that: 1) With fixed $G_V$ and fixed $a_0$, the transverse
pressure density for $B_0=4\times 10^{18}G$ is higher than that for $B_0=0$; 2) With
fixed $G_V$, the positive magnetized gluon
pressure density ($a_0=0.01$) and the case $B_0 =4\times 10^{18}G$ always gives the
hardest equation of state, while the negative magnetized gluon pressure ($a_0=-0.01$)
and the case $B_0 =0$ always gives
the softest equation of state; 3) For the case of negative magnetized gluon pressure ($a_0=-0.01$),
it is found that the equations of state for both $B_0=0$ and $B_0=4\times 10^{18}G$ are not
sensitive to the value of $G_V$. However, for the case of positive magnetized gluon pressure ($a_0=0.01$), the EoS for both $B_0=0$ and $B_0=4\times 10^{18}G$ are very
sensitive to the value of $G_V$; 4)When $G_V=0$, the magnetic field contribution to equation
of state is important, while when $G_V$ increases, the contribution from magnetized
gluon pressure becomes more and more important to the equation of state.

Shown in Fig.~\ref{PLE} is the longitudinal pressure as a function of energy density
of SQM in cases of $B_0=0$ and $B_0=4\times 10^{18}G$
with $a_0=0,0.01,-0.01$ and $G_V=0$, $G_V=0.4G_S$, $G_V=0.8G_S$ and $G_V=1.1 G_S$,
respectively. From this figure one can find that: 1) For $G_V=0$ and fixed $a_0$,
the longitudinal pressure density for $B_0=4\times 10^{18}G$ is a little
smaller than that at $B_0=0$, which is opposite to the case of transverse pressure;
2) With fixed $G_V$, the positive magnetized gluon
pressure ($a_0=0.01$) and the case $B_0 =0$ always gives the hardest equation of
state, while the negative magnetized gluon pressure ($a_0=-0.01$) and the case
$B_0 =4\times 10^{18}G$ always gives the softest equation of state;
3) For the case of negative magnetized gluon pressure ($a_0=-0.01$),
it is found that the equations of state for both $B_0=0$ and $B_0=4\times 10^{18}G$
are not sensitive to the value of $G_V$. However, for the case of positive magnetized
gluon pressure ($a_0=0.01$), the equations of state for both $B_0=0$ and
$B_0=4\times 10^{18}G$ are very sensitive to the value of $G_V$. Compared to
Fig~\ref{PE}, we can find that the pressure anisotropy for longitudinal and
transverse pressure is small when $G_V$ is as big as $G_V=0.8G_S$, $G_V=1.1 G_S$ for
$a_0=0.01$, while the pressure anisotropy gets larger as the decrement of $G_V$ for
$a_0=-0.01$. Compared to the result from Fig.~\ref{PE}, we find the magnetized gluon
pressure contribution is more important to stiffen the EoS for SQM than the contribution
from magnetic field, and by using this contribution from magnetized gluon one can
describe a heavy QS (about 2$M_{\odot}$) with small pressure anisotropy
(like $G_V=0.8G_S$, $G_V=1.1 G_S$ for $a_0=0.01$ cases) under a reasonable
magnetic field distribution inside QSs.

\begin{figure}[tbp]
\includegraphics[scale=0.42]{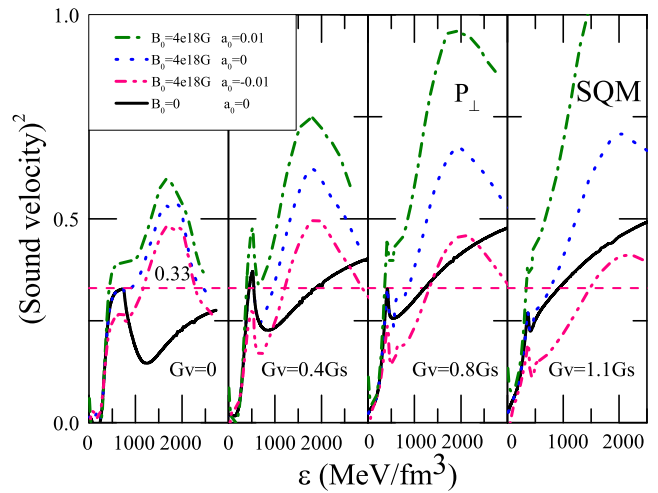}
\caption{(Color online) The sound velocity square for transverse pressure as a function of energy density for SQM in cases of $B_0=0$ and $B_0=4\times 10^{18}G$
with $a_0=0,0.01,-0.01$ and $G_V=0$, $G_V=0.8G_S$ and $G_V=1.1 G_S$,
respectively.}
\label{Fig:cs}
\end{figure}

\begin{figure}[tbp]
\includegraphics[scale=0.42]{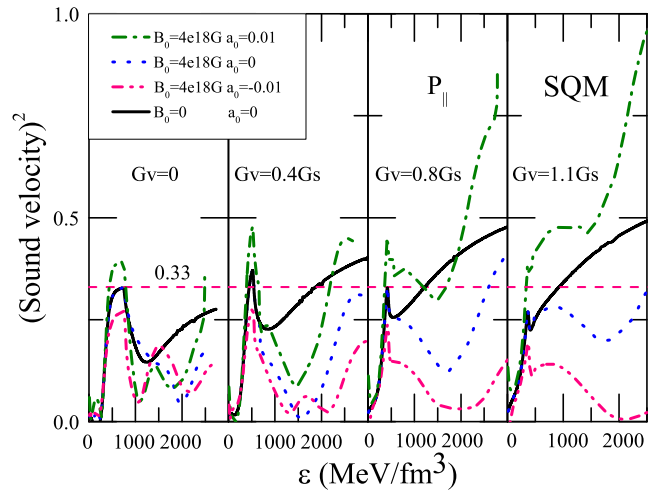}
\caption{(Color online) The sound velocity square for longitudinal pressure as a function of energy density for SQM in cases of $B_0=0$ and $B_0=4\times 10^{18}G$
with $a_0=0,0.01,-0.01$ and $G_V=0$, $G_V=0.8G_S$ and $G_V=1.1 G_S$,
respectively.}
\label{Fig:cs2}
\end{figure}

It was pointed out in the work \cite{Alford:2013aca} that to construct a hybrid star with mass
heavier than $2 M_\odot$, a large sound velocity square for quark matter, say larger than 1/3
is preferred. It is known that the sound velocity square for ideal gas or for
strongly coupled conformal theory can be $1/3$. However, for strongly interacting liquid,
the sound velocity square is normally smaller than $1/3$. Therefore, it is interesting to
investigate the sound velocity in our current model, and the results of sound velocity
\begin{equation}
c_s^2=\frac{d p}{d \epsilon}
\end{equation}
directly derived from equations of state in Fig.~\ref{PE} and Fig.~\ref{PLE} are shown in Fig.\ref{Fig:cs} and Fig.\ref{Fig:cs2}. Because the main purpose of this work is to explore the properties of SQM under strong magnetic field by considering the magnetized gluon contribution, we choose $B_0=4\times 10^{18}G$
with $a_0=0,0.01,-0.01$ and $B_0=0$ with $a=0$ by considering the vector-isoscalar interaction as $G_V=0$, $G_V=0.8G_S$ and $G_V=1.1 G_S$ in the following parts.


The sound velocity square of dense quark matter for transverse pressure density
in Fig~\ref{Fig:cs} in the NJL model
without vector interaction ($G_V=0$) and at $B_0=0$ is always
less than 1/3, which is in agreement with common sense. It is also understood that
when the system is more strongly coupled, the sound velocity square is getting smaller,
and reach the smallest value at the phase transition point, where the system is regarded
as most strongly coupled \cite{Mao:2009aq}.

However, it is observed that under magnetic field with magnitude of $B_0= 4\times 10^{18}G$,
even at $G_V=0$, the sound velocity square for transverse pressure density case can increase
to 0.6 at most. Also in the case of $B_0=0$, if one switches on a repulsive interaction in the vector-isoscalar channel,
the sound velocity square for transverse pressure density case also increases and
can become bigger than $1/3$. The stronger the
$G_V$ is, the larger sound velocity square will be.

Another factor to increase the sound velocity is from the magnetized gluons. With a positive
contribution from magnetized gluon pressure, and also taken into account the repulsive
interaction in the vector-isoscalar channel, the sound velocity for transverse pressure case can be even larger than 1(for $G_V=1.1G_S$, $B_0=4\times10^{18}G$ and $a_0=0.01$ case),
i.e., larger than the speed of light, which is of course not physical. So we can use the
condition $c_s^2<1$ to constrain the EoS.

For the sound velocity square corresponding to longitudinal pressure from Fig.\ref{Fig:cs2},
we can find that the sound velocity square under $B_0= 4\times 10^{18}G$ with different
$G_V$ and $a_0$ are all smaller than those $C_s^2$ corresponding to the transverse pressure
case, which is self-consisted with Fig.~\ref{PE} and Fig.~\ref{PLE}. One can also
find that all the sound velocity square corresponding to longitudinal pressure
are smaller than 1.

Since we have calculated that the pressure anisotropy for SQM with the contribution
from magnetized gluon pressure under the magnetic field distribution inside the QSs within
$B_0=4\times10^{18}G,~\beta_0=0.003~\text{and}~\gamma=3$ is not big, we can approximately
use the EoS from longitudinal or transverse pressure case to construct the magnetar under
density dependent magnetic field. We should firstly introduce the TOV equations
\citep{Oppenheimer39} which can give the quark star with isotropic
pressure:
\begin{align}
\frac{dM}{dr}=4\pi r^2 \epsilon(r),
\end{align}
\begin{eqnarray}
\frac{dp}{dr}&=&-\frac{G\epsilon(r)M(r)}{r^2}[1+\frac{p(r)}{\epsilon(r)}]\nonumber \\
& & [1+\frac{4\pi p(r)r^3}{M(r)}][1-\frac{2GM(r)}{r}]^{-1}.
\end{eqnarray}

\begin{figure}[tbp]
\includegraphics[scale=0.36]{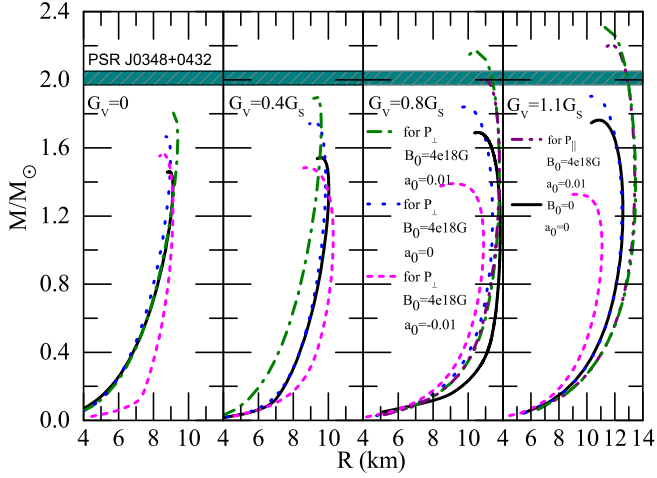}
\caption{(Color online)Maximum mass - radius relation for quark star for transverse and longitudinal pressure cases within $B_0=0$
and $B_0=4\times 10^{18}G$ with $a_0=0,0.01,-0.01$ and $G_V=0$, $G_V=0.4G_S$, $G_V=0.8G_S$
and $G_V=1.1 G_S$, respectively. }
\label{RMGIV}
\end{figure}

We give the maximum mass of quark star in Figure~\ref{RMGIV} by using transverse pressure and longitudinal pressure respectively, and we use $B_0=4\times 10^{18}G$ with $a_0=0,0.01,-0.01$ and $G_V=0$, $G_V=0.4G_S$, $G_V=0.8G_S$ and $G_V=1.1 G_S$ for transverse pressure case while $B_0=4\times 10^{18}G$, $a_0=0.01$ with $G_V=0.8G_S$ and $G_V=1.1 G_S$ for longitudinal pressure case.
We can read the following information from Figure~\ref{RMGIV}: 1) At $B_0=0$ and $G_V=0$, the three-flavor NJL model
give the maximum mass of quark star is about $1.4 M_{\odot}$; 2) At
$B_0=0$ but increases the repulsive interaction $G_V$ in the vector-isoscalar channel, the
the maximum mass of quark star can reach $1.75 M_{\odot}$ for $G_V=1.1 G_S$; 3) In the case
of $G_V=0$, when put quark matter under the magnetic field, the maximum mass of quark
magnetar for transverse pressure case
can be as heavy as $1.65 M_{\odot}$, and if one takes into account a positive
pressure density
contributed from magnetized quasi-gluons, the mass of quark magnetar can reach $1.8 M_{\odot}$;
4) For the magnitude of $B_0=4\times 10^{18}G$,
the sound velocity square for transverse pressure case reaches almost 0.9 for $G_V=0.8 G_S$ and $a_0=0.01$, and the magnetar mass is $2.17 M_{\odot}$, while the maximum mass of quark star is $2.01M_\odot$ for longitudinal pressure within $G_V=0.8 G_S$ and $a_0=0.01$ under $B_0=4\times 10^{18}G$, which is consistent with the recently discovered large mass pulsar J0348+0432 $(2.01\pm0.04)M_{\odot}$.

In order to investigate the difference for the maximum mass of QSs by using longitudinal
pressure and transverse pressure, we define a mass difference parameter
\begin{eqnarray}
\delta_m=\frac{M_{\perp}-M_{||}}{(M_{\perp}+M_{||})/2},
\end{eqnarray}
where $M_{\perp}$ ($M_{||}$) represents the maximum mass of QSs with
transverse (longitudinal) orientation pressure, respectively.
We calculate the mass difference for the transverse and longitudinal pressure
within $G_V=0.8 G_S$ and $a_0=0.01$ under $B_0=4\times 10^{18}G$ is $\delta_m=7.65\%$,
which implies that the pressure anisotropy in this case is very small due to the tiny mass asymmetry, and this is the reason why we can use the isotropic TOV equation to calculate the properties of QSs approximately. We can also find the similar results from $G_V=1.1 G_S$, $a_0=0.01$ and $B_0=4\times 10^{18}G$ case that the maximum mass of QS for transverse pressure case is $2.30M_\odot$ while $2.20M_\odot$ for longitudinal pressure case, and the mass difference for this case is $\delta_m=4.44\%$, which implies a smaller pressure anisotropy than $G_V=0.8G_S$ case. Therefore, our results indicate that we can get stiffer EoS by considering the contributions from density dependent magnetic field, the repulsive interaction in the vector-isoscalar channel and magnetized gluon pressure. Since the pressure anisotropy from density-dependent magnetic field is not big, we can calculate the properties of QSs under magnetic field by using isotropic TOV equation approximately, and we find the mass difference of magnetars by using transverse and longitudinal
pressure is also very small.

\begin{figure}[tbp]
\includegraphics[scale=0.41]{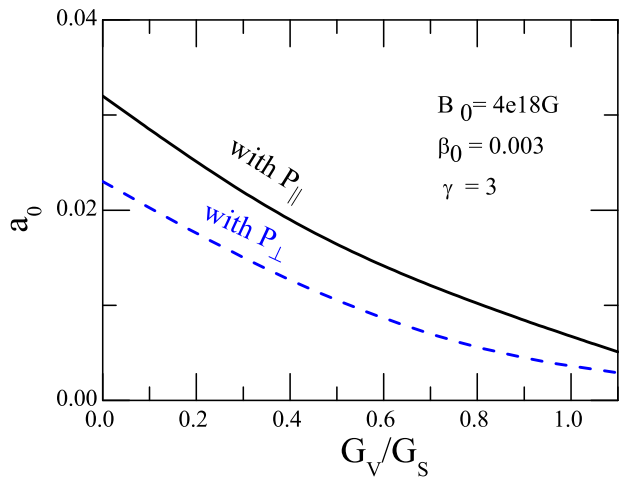}
\caption{(Color online)The parameter region for $a_0$ and $G_V$ to produce 2 solar mass QSs under $B_0=4\times 10^{18}G$ and by using longitudinal pressure and transverse pressure, respectively. The parameter sets of $a_0$ and $G_V$ on this two lines can produce 2 solar
mass QSs.}
\label{GVa0}
\end{figure}

In Fig.~\ref{GVa0}, we show the parameter region for $a_0$ and $G_V$ to produce 2 solar mass QSs under $B_0=4\times 10^{18}G$ and by using longitudinal pressure and transverse pressure, respectively. The parameter sets of $a_0$ and $G_V$ on this two lines can describe 2 solar
mass QSs. One can see that in order to produce 2 solar mass QSs, if the repulsive vector
interaction is stronger, the needed contribution from magnetized gluons is smaller, or vice versa.
It is also observed that larger $a_0$ and $G_V$ parameters are needed to produce 2 solar mass QSs
if one use the longitudinal pressure density. However, when the magnitude of repulsive vector
interaction increases, it is noticed that the difference between using longitudinal pressure and transverse pressure to produce 2 solar mass QSs becomes smaller, because the pressure anisotropy
decreases when large contribution from the repulsive interaction in the vector-isoscalar channel
is considered, which can be read from Fig.~\ref{PE} and Fig.~\ref{PLE}.

\section{conclusion and discussion}

In this work, we construct quark magnetars in the framework of SU(3)
NJL model with vector interaction under strong magnetic field.

We investigate the effect of vector-isoscalar and vector-isovector
interaction on the equation of state, and it is found that the equation of
state is not sensitive to the vector-isovector interaction, however, a repulsive
interaction in the vector-isoscalar channel gives a stiffer equation
of state for cold dense quark matter. The result is reasonable because
the interaction in the vector-isovector channel shifts the isospin
chemical potential, and this effect is negligible on the equation of state
under $\beta$-equilibrium, while the repulsive vector-isoscalar interaction shifts
the chemical potential to a larger value, which makes the equation of state stiffer.

In the presence of magnetic field, the pressure of the system is shown to be anisotropic
along and perpendicular to the magnetic field direction
with the former being generally larger than the latter. Gluons will be magnetized via
quark loops, and we also estimate the pressure contributed from magnetized gluons.
Normally the NJL model only considers the contribution from quark degrees of freedom
on the pressure, which is always underestimated. We estimate the pressure density contributed
from magnetized gluons, which should be positive in order to compensate the pressure
density of quasi-quarks described by the NJL model. It is found that magnetized quarks
and gluons also give a stiffer equation of state.

The sound velocity square is one of fundamental properties of hot/dense matter, which
measures the hardness or softness of the equation of state. It is
known that hot and dense quark matter in the NJL model without vector interaction at
zero magnetic field is always less than 1/3. It is also understood that
when the system is more strongly coupled, the sound velocity square is getting smaller,
and reach the smallest value at the phase transition point, where the system is regarded
as most strongly coupled. However, it is found that the sound velocity square can be larger
than 1/3 when a strong repulsive interaction is introduced, and the sound velocity square
can change a lot when strong magnetic field is added. Furthermore, the sound velocity
square corresponding to the transverse pressure density case can even reach 1 when the
magnetized gluon contribution is taken into account.

Since the pressure anisotropy from density-dependent magnetic field is small, we can
calculate the properties of QSs under magnetic field by using isotropic TOV equation
approximately, and we find the magnetar mass difference by using transverse and longitudinal
pressure cases are very small and the maximum mass of quark star can be enhanced by the
contribution from the density-dependent magnetic field, the repulsive interaction in
the vector-isovector channel and the magnetized gluon pressure. We also give
the parameter region for $a_0$ and $G_V$ which can describe the 2 solar mass
quark star by using longitudinal pressure and transverse pressure respectively.

\vskip 1 cm
{\bf Acknowledgement.---} We thank valuable discussions with M. Alford and
A. Schmitt on sound velocity of quark matter, with J. Schaffner-Bielich and A. Sedrakian
on magnetars, and with I. Shovkovy and A. Vuorinen on pressure from magnetized gluon
potential. M. Huang thanks the hospitality of Frankfurt University, Tours University and TU Viena, where
the final stage of this work is performed. This work is supported by the National Basic Research Program of China (973 Program) under Contract
No. 2015CB856904 and 2013CB834405, the NSFC under Grant
Nos. 11275213, 11275125, 11135011 and 11261130311(CRC 110 by DFG and NSFC), also
supported by CAS key project KJCX2-EW-N01, and Youth Innovation Promotion Association
of CAS, the China-France collaboration Project "Cai Yuanpei 2013", the Shanghai
Rising-Star Program under grant No. 11QH1401100, the ``Shu Guang" project supported
by Shanghai Municipal Education Commission, and Shanghai Education Development Foundation,
the Program for Professor of Special Appointment (Eastern Scholar) at Shanghai Institutions
of Higher Learning, and the Science and Technology Commission of Shanghai Municipality (11DZ2260700).

\end{document}